\documentclass[epj]{svjour}
\usepackage{psfig,epsfig}
\usepackage{graphicx}
\usepackage{float,pst-all}

\begin{document}
\title{Role of Higher Multipole Excitations in the Electromagnetic Dissociation of One Neutron Halo Nuclei}
%\subtitle{Do you have a subtitle?\\ If so, write it here}
\author{R. Chatterjee{\inst{1,2}}, L. Fortunato\inst{1} \and A. Vitturi\inst{1}% etc
 %\thanks is optional - remove next line if not needed
%\thanks{\emph{$^a$rajdeep.chatterjee@pd.infn.it,$^b$lorenzo.fortunato@pd.infn.it, $^c$andrea.vitturi@pd.infn.it } }%
}                     % Do not remove
%
%\offprints{}          % Insert a name or remove this line
%
\institute{Dipartimento di Fisica and INFN, Universit\`a di Padova, via F. Marzolo 8, I-35131, Padova, Italy 
\and Theory Group, Saha Institute of Nuclear Physics, 1/AF, Bidhannagar, Kolkata 700064, India}
\date{Received: date / Revised version: date}
% The correct dates will be entered by Springer
%
\abstract{
We investigate the role of higher multipole excitations in the electromagnetic dissociation of one-neutron halo nuclei
within two different theoretical models -- a finite range distorted wave Born approximation and another in a more analytical
method with a finite range potential. We also show, within a simple picture, how the presence of a weakly bound state affects the breakup
cross section.
\PACS{
      {24.10.-i}{Nuclear reaction models and methods}   \and
      {24.50.+g}{Direct reactions}
     } % end of PACS codes
} %end of abstract
\authorrunning{R. Chatterjee, L. Fortunato, A. Vitturi}
\titlerunning{Higher Multipole Excitations in Electromagnetic Dissociation ..}
\maketitle

%%%%%%%%%%%%%%%%%%%%%%%%%%%%%% INTRODUCTION  %%%%%%%%%%%%%%%%%%%%%%%%%%%%%%%
%%%%%%%%%%%%%%%%%%%%%%%%%%%%%%%%%%%%%%%%%%%%%%%%%%%%%%%%%%%%%%%%%%%%%%%%%%%%

\section{Introduction}
\label{intro}
%(Direct reactions with exotic beams:)
%Coulomb excitation and dissociation are established methods to study a wide variety of nuclei
%from stable to weakly bound drip line nuclei. 
The electromagnetic or Coulomb dissociation method is a well established method to study
the properties of a wide variety of nuclei -- from stable to weakly bound drip line ones. That the
electromagnetic interaction is quantifiable has led to the development of several analytical 
and semi-analytical reaction models in
nuclear physics ranging from the semi-classical to the purely quantal.
Theoretical studies in Coulomb breakup reactions have also received a boost, in recent years, with increasing
interest in nuclei far from the valley of stability, where one encounters isotopes which
are often extremely unstable (especially those closer to the driplines) 
and have structure and properties  different from stable ones.
Investigating these nuclei also opens up the interesting prospect of testing the limits
of theoretical models across the nuclear chart.

Coulomb dissociation has a rich history of 
being used as a probe to investigate projectile structure information. For instance, it would place constraints 
on their electric dipole response \cite{nak1,nak2,ber88,ber91}. 
It has also been used as an useful indirect method in nuclear
astrophysics. One can relate the Coulomb breakup cross section to the corresponding photodisintegration
cross section and in turn relate it to the inverse radiative capture cross section \cite{bbh}. One is thus able to
simulate and measure reaction cross sections, in the laboratory on earth, of stellar reactions which goes on at extremely
low energies. This method thus provides an ideal theoretical laboratory to study the physics of breakup reactions as 
a tool for nuclear structure and astrophysics \cite{bht}.

In this paper we primarily investigate the role of higher multipole excitations as a function of
neutron separation energy in the electromagnetic dissociation of one neutron halo nuclei with direct
reactions and with two different theoretical methods. The first method is based on the post form finite range distorted wave Born 
approximation (FRDWBA) \cite{chatt} (henceforth referred to as Method 1, see section 2.1). In this model the electromagnetic 
interaction between the core and the target nucleus is included to all orders and the breakup contributions from the
entire non-resonant continuum corresponding to all the multipoles 
and the relative orbital angular momenta between the valence nucleon and the core fragment are included.
 Full ground state wave function of the
projectile, of any angular momentum configuration, enters as an
input to the theory. This method has also been referred to as the Coulomb wave Born approximation
in the literature \cite{pb}. The results obtained within this method are compared with those obtained from a 
standard first order Coulomb dissociation theory based on multipole expansion under the aegis of the Alder-Winther theory \cite{ald} 
(henceforth indicated with Method 2, see section 2.2). 
The multipole strengths entering in this second approach can be calculated in a simple single-particle picture for one-neutron halo nuclei 
as a function of different one-neutron separation energies (artificially varied from very weakly bound to more stable systems).
For the single-particle potential describing the interaction between the valence(halo) neutron and the core, we will either use 
a Woods-Saxon potential, which has to be numerically solved, or a square well potential, leading to more analytical expressions \cite{nlv,bt}
for the $B(E\lambda)$ distributions.
 Our aim is to see how far the predictions from these two models tally with each other.

The comparison of the two methods allows us to address the relative importance of dipole and quadrupole breakup contributions in 
Coulomb dissociation.
We shall show the sensitivity of the reaction observables to the details of nuclear structure: in particular, in the 
case of $^{11}$Be, the low-lying continuum is strongly affected by the presence of a weakly-bound $1p_{1/2}$ excited state.

%%  CUT, CHANGE or KEEP at your leisure. Seems redundant after our changes.
%
%Thus, in this paper, we compare different theoretical models of breakup reactions by calculating several reaction observables like  
%relative energy spectra, angular distributions and one neutron removal cross section, taking the neutron separation energy as a parameter.
%In a way, we theoretically simulate weakly bound isotopes to stable ones for the same angular momentum configuration of the system, which in 
%turn opens up the possibility to study the applicability of different theoretical models in a wide variety of situations: from the valley of %stability to the neutron drip line.

The paper is organised in the following way. A brief summary of the FRDWBA formalism and the relevant analytical quantal results with 
the finite range potential 
are given in section {\bf 2}. Our results, with a description of the structure model, relevant reaction cross sections
and the effect of having a weakly bound state near the particle emission threshold are discussed in section {\bf 3}.
 Summary and conclusions of our work are in section {\bf 4}.

%%%%%%%%%%%%%%%%%%%%%%%%%%%%%%% FORMALISM  %%%%%%%%%%%%%%%%%%%%%%%%%%%%%%%
%%%%%%%%%%%%%%%%%%%%%%%%%%%%%%%%%%%%%%%%%%%%%%%%%%%%%%%%%%%%%%%%%%%%%%%%%%
\section{Formalism}
\label{sec:1}
\subsection{Finite range distorted wave born approximation - Method 1}
We consider the reaction $ a + t \rightarrow b + c + t $, where the 
projectile $a$ breaks up into fragments $b$ (charged) 
and $c$ (uncharged) in the Coulomb
field of a target $t$. 
The triple
differential cross section for the reaction is given by 
\begin{eqnarray}
{d^3\sigma \over dE_b d\Omega _b d\Omega _c}={2\pi\over \hbar v_a}\left\{\sum_{l
\mu}\frac{1}{(2l + 1)}\vert \beta_{l\mu}\vert^2 \right\} \rho(E_b,
\Omega _b,\Omega _c)~.
\end{eqnarray}
Here $v_a$ is the $a$--$t$ relative velocity in the entrance channel and
$\rho (E_b,\Omega _b,\Omega _c)$ the phase space factor appropriate 
to the three-body final state \cite{fuchs}. The reduced amplitude $\beta_{l\mu}$
in post form finite range distorted wave Born approximation is given by
\begin{eqnarray}
\beta_{l\mu}
=& & \langle \exp(\gamma\vec{k}_c-\alpha\vec{K})\vert V_{bc}\vert \Phi _{a}^{l\mu}\rangle \nonumber \\
&\times &  \langle \chi ^{(-)}(\vec{k}_b)\chi^{(-)}(\delta\vec{k}_c)\vert \chi^{(+)}(\vec{k}_a) \rangle~, \label{dw}
\end{eqnarray}
where, $\vec{k}_b$, $\vec{k}_c$ are Jacobi wave vectors of fragments $b$
and $c$, respectively in the final channel of the reaction, $\vec{k}_a$ is the wave vector of
projectile $a$ in the initial channel and $V_{bc}$ is the 
interaction between $b$ and $c$. $\Phi _{a}^{l\mu}$ is the ground state wave function
of the projectile with relative orbital angular momentum state $l$ and projection $\mu$.
In the above, $\vec{K}$ is an effective local momentum associated with the
core-target relative system, whose direction has been taken to be the same as 
the direction of the asymptotic momentum $\vec{k}_b$ \cite{nag,chatt}. $\alpha, \delta$ and $\gamma$ in Eq. 2, are 
mass factors relevant to the Jacobi coordinates of the three body system (see Fig. 1 of Ref. \cite{chatt}).
$\chi^{(-)}$'s are the distorted waves for relative motions of 
$b$ and $c$ with respect to $t$ and the center of mass (c.m.) of the $b-t$
system, respectively, with ingoing wave boundary condition and $\chi^{(+)}(\vec{k}_a)$ is
the distorted wave for the scattering of the c.m. of projectile $a$ with respect to the target 
with outgoing wave boundary condition.

Physically, the first term in Eq. (\ref{dw}) contains the structure information about
the projectile through the ground state wave function $\Phi _{a}^{l\mu}$, and is known as the 
vertex function, while the second term is associated only with the dynamics of the reaction.
The charged projectile $a$ and the fragment $b$ interacts with the target by a point Coulomb 
interaction and hence $\chi^{(-)}_b(\vec{k}_b)$ and $\chi^{(+)}(\vec{k}_a)$ are substituted with
appropriate Coulomb distorted waves.
For pure Coulomb breakup, of course, the interaction between
the target and uncharged fragment $c$ is zero and hence
$\chi^{(-)}(\delta\vec{k}_c)$ is replaced by a plane wave. This will allow the
second term of Eq. (\ref{dw}), the dynamical part, to be evaluated analytically in terms
of the bremsstrahlung integral \cite{nor}.
  
A more detailed description of the formalism can be found in Refs. \cite{chatt,pb}.

\subsection{First order multipole Coulomb dissociation -Method 2}
%An analytic quantal model with a finite range potential

Alternative to the previous method, more standard approaches are based on the multipole 
expansion of the Coulomb field. In the time-dependent Alder-Winther formalism, adapted for
continuum states, the excitation probability for a given impact parameter and bombarding 
energy is proportional to the $dB(E\lambda)/dE_{bc}$ distribution. In the case of high bombarding
energies the kinematic part of the dipole cross-section can be interpreted in terms of equivalent 
photon number, $n_{E1}$, leading to
\begin{equation} 
{d\sigma_{E1} \over {dE_{bc}}} = {dB(E1) \over {dE_{bc}}} n_{E1}.
\end{equation}
If one further assumes that $n_{E1}$ is weakly dependent on
the fragment -- fragment relative energy after breakup, then the breakup cross-section $\sigma_{E1}$ is
directly proportional to the total B(E1). 

Within this line of reasoning all the information on the structure is contained in the matrix element
of the electromagnetic operator. In a single-particle description of the dissociation of a halo nucleus, 
the transition can be attributed to the promotion of the valence neutron from a bound to a continuum state.
The $B(E\lambda)$ is obtained by evaluating the matrix element involving the initial bound and final continuum 
wave functions defined in the projectile mean-field potential.

For the sake of pushing the mathematical treatment as far as possible one 
can choose a finite square well potential \cite{nlv}.
In the limit of small binding energy, the bound and unbound states may be written in terms of their asymptotic 
form as first order spherical Henkel functions and spherical Bessel functions 
(of an appropriate order) respectively. Since the energy is small, 
the largest part of the 
contribution to the dipole strength will come from the outer region. This 
fact allows one to derive simple expressions for the strength distribution,
namely $dB(E1)/dE_{bc}$, in terms of the binding energy, $S_n$, and the relative continuum 
energy, $E_{rel} = E_{bc}$. As an explicit example, if one considers electric transitions 
from a weakly bound $s$-orbit to the $p$-continuum, the dipole strength is given by
\begin{equation}
dB(E1)/dE_{bc} \propto {\sqrt{S_n}E_{bc}^{3/2} \over (E_{bc}+S_n)^4}~.
\label{dip}
\end{equation}
This distribution has a maximum for $E_{bc}=3/5S_n$ \cite{nlv}. This 
prediction will be discussed later in the paper.

As found in Ref. \cite{nlv,bt} in this extreme 
single-particle picture the total integrated $B(E1)$ is connected to the 
mean square radius, $r$, of the single-particle state, namely \cite{nlv}:
\begin{equation}
B(E1)=(Z_{eff}^{(1)} e)^2 {3\over 4\pi} \langle 
r^{2} \rangle
\end{equation}
and more in general for any multipolarity, $\lambda$, (Eq. 65 in Ref.\cite{bt})
\begin{equation}
B(E\lambda)=(Z_{eff}^{(\lambda)} e)^2 {2\lambda+1\over 4\pi} \langle 
r^{2\lambda} \rangle
\end{equation}
where the effective charge is defined as
\begin{equation}
Z_{eff}^{(\lambda)}=Z_c\biggl({A_b\over A}\biggr)^\lambda +
Z_b\biggl({-A_c\over A}\biggr)^\lambda ~,
\end{equation}
with $Z_i$ and $A_i$ $(i=b,c)$ being the charges and atomic numbers of breakup fragments, respectively
and $A$ being the atomic number of the projectile.

If the potential is a square well the expectation value of the $r^{2\lambda}$
operator in the ground state $s$ wave function is:
\begin{equation}
\langle r^{2\lambda} \rangle ={(2\lambda)!\over (2a)^{2\lambda}}
\end{equation}
and the reduced transition probability becomes
\begin{equation} \label{be1}
B(E\lambda)=(Z_{eff}^{(\lambda)} e)^2 {(2\lambda+1)!\over 4\pi(2a)^{2\lambda}}
= (Z_{eff}^{(\lambda)} e)^2 {(2\lambda+1)!\over 4\pi}
\Biggl({\hbar^2\over 8\mu S_n}\Biggr)^\lambda
\end{equation}
having used the relation $a^2=2\mu S_n/\hbar^2$,  with $\mu$ being the reduced mass of the fragment -- fragment
system in the final channel.

 Therefore in the present case by using Eq. (\ref{be1}), one obtains
\begin{equation} \label{lf3}
\sigma_{E1} \propto B(E1) \propto {1\over S_n}.
\end{equation}
that relates the total dipole cross-section with the one-neutron separation energy.

%This simple estimate for the total integrated multipole strengths suggests 
%a relationship with the mean values of the $r^{2\lambda}$ operator. 
%Therefore, when considering more sophisticated models, one could check 
%whether the evaluation of one-neutron break-up cross-section (which is 
%proportional to the sum of the various multipole contributions) does yield 
%a direct proportionality to $\langle r^2\rangle$. If it doesn't one should 
%look for higher order effects as a possible cause. 

%%%%%%%%%%%%%%%%%%%%%% FRDWBA & FINITE DEPTH WELL RESULTS  %%%%%%%%%%%%%%%%%%%%%%%%%%
%%%%%%%%%%%%%%%%%%%%%%%%%%%%%%%%%%%%%%%%%%%%%%%%%%%%%%%%%%%%%%%%%%%%%%%%%%%%

\section{Results}
\subsection{Ground state structure of $^{11}$Be and $^{19}$C}

To describe the structure of one-neutron halo nuclei $^{11}$Be and $^{19}$C, we use single particle wave functions
for the valence neutron constructed by assuming the neutron-core 
interaction to be of Woods-Saxon type. For the purpose of better understanding the
properties of weakly bound nuclei, we artificially vary the depth of the potential in order to 
obtain a set of different binding energies (among which also the experimental one). In this way 
we simulate situations ranging from nuclei at the valley of stability to the neutron drip line for the
same angular momentum configuration.

For the ground state of $^{11}$Be, we have considered the following 
configuration : a $s$ -- wave valence neutron coupled to $0^+$  
$^{10}$Be core, namely [$^{10}$Be$(0^+) \otimes 1s_{1/2}\nu$] with a   
one-neutron separation energy, $S_{n}$.
A similar method is adopted for $^{19}$C : a  $^{18}$C  $(0^+)$ core coupled to a neutron in the $2s$ orbital. 
 The radius and diffuseness parameters of the Wood-Saxon well for each case have been taken to be  1.15 fm and 0.5 fm,
respectively.
  The list of parameters
for the two isotopes is given in table 1.

\begin{table}[ht]
\caption{Potential parameters. The single particle wave function,  
is constructed by assuming the valence neutron-core
interaction to be of Woods-Saxon type whose depth is adjusted to reproduce
the corresponding value of the binding energy with fixed values of the radius
and diffuseness parameters (taken to be  1.15 fm and 0.5 fm,
respectively)}
\begin{center}
\begin{tabular}{|c|c|c|}
\hline
 Projectile &  $S_n$ & $V_{depth}$  \\
 & (\footnotesize{MeV})& (\footnotesize{MeV}) \\ 
\hline
$^{11}$Be &  0.100 & 67.33  \\
         &  0.250 & 69.00  \\
         &  0.504 & 70.99  \\
         &  0.750 & 72.53  \\
         &  1.000 & 73.89  \\         
         & ~      &    \\
 $^{19}$C  & 0.160 & 47.43  \\
           &0.350 & 48.77  \\
           & 0.530 & 49.77  \\
           &0.750 & 50.82  \\
           & 1.000 & 51.86  \\               
\hline
\end{tabular}
\end{center}
\end{table}

\subsection{Relative energy spectra and angular distributions within the FRDWBA}
In this subsection we shall calculate within the FRDWBA (method 1) the relative energy spectra in the breakup of $^{11}$Be
and $^{19}$C on a heavy target for different binding energies of the projectile 
by varying the one neutron separation energy, for the same angular momentum configuration of the system.
We shall also present the angular distribution of the
projectile c.m. with respect to the target in the breakup of $^{11}$Be on $^{208}$Pb
at four different beam energies.

\begin{figure}
\centerline{{\epsfig{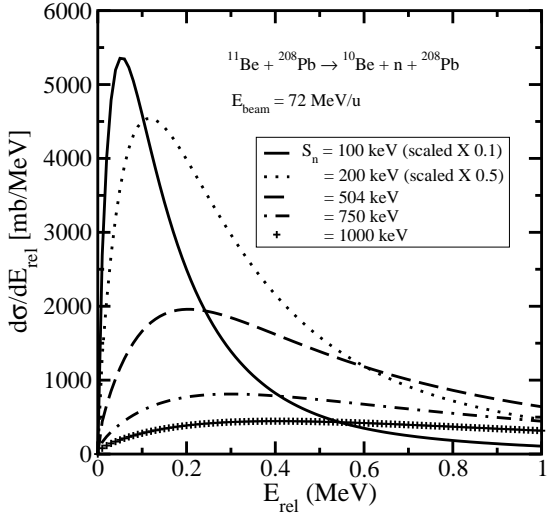}}}
% Use the relevant command for your figure-insertion program
% to insert the figure file.
% For example, with the option graphics use
%\resizebox{0.35\textwidth}{!}{%
%  \includegraphics{rel_be.eps}
%}
% If not, use
%\vspace{5cm}       % Give the correct figure height in cm
\caption{Relative energy spectra in the breakup of $^{11}$Be 
on a $^{208}$Pb target at 72 MeV/u incident beam energy, for different 
one-neutron separation energies. For $S_n = 100 \textrm{ and } 200$ keV
the results are scaled by 0.1 and 0.5, respectively.}
\label{fig1}       % Give a unique label
\end{figure}

In  Fig. \ref{fig1}, we show the relative energy spectra for the breakup of $^{11}$Be 
on a $^{208}$Pb target at 72 MeV/u incident beam energy, for different 
one-neutron separation energies: $S_n$ = 100 keV (solid), 200 keV (dotted),
504 keV (dashed), 750 keV (dot-dashed) and 1000 keV (plus signs). Among them
the actual experimental value is 504 keV.
Note that for $S_n = 100 \textrm{ and } 200$ keV
the results are scaled by multiplying them with 0.1 and 0.5, respectively.

\begin{figure}
\centerline{{\epsfig{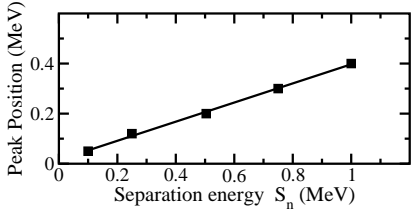}}}
%\resizebox{0.25\textwidth}{!}{%
%  \includegraphics{rel_peak.eps}
%}
\caption{\label{figb1v}Peak of the relative energy spectra as a function
of (fictitious) one-neutron separation energies 
in the breakup of $^{11}$Be on a $^{208}$Pb target at
72 MeV/nucleon.}
\end{figure}

It is to be noted that the bound state neutron was in the $s$-state (initial
channel) and the transition to all possible states in the final channel has been
taken care of by the post form of the transition matrix (Eq. 2). 
To compare our results with those predicted from the analytical model \cite{nlv},
we plot the peak of the relative energy spectra as a function
of one-neutron separation energy in the breakup of $^{11}$Be on a $^{208}$Pb target at
72 MeV/nucleon in Fig. \ref{figb1v}. 

It is seen that the slope is approximately 0.4 which is pretty close to the
simple estimate obtained from distribution \ref{dip}.
The discrepancy can be attributed to the approximate description of the wave functions
implicit in the derivation of the formula.

Fig. \ref{figb1v} could also serve a different purpose. Since the dependence of the peak of the
relative energy spectra as a function of the one-neutron separation energy is approximately linear, it could also be used
to get an heuristic idea about the separation energy of a system from the experimental relative energy spectra. 
\begin{figure}
\centerline{{\epsfig{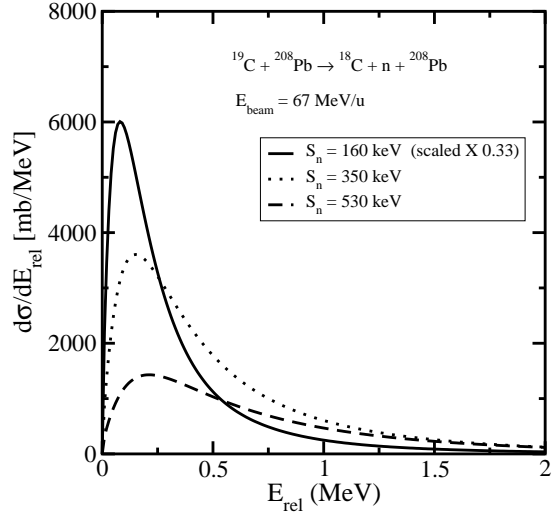}}}
%\resizebox{0.4\textwidth}{!}{%
%  \includegraphics{rel_c67.eps}}
\caption{\label{figa1c}Relative energy spectra in the breakup of $^{19}$C 
on a $^{208}$Pb target at 67 MeV/u incident beam energy, for different 
(fictitious) one-neutron separation energies. For $S_n$ = 160 keV
the results are scaled 0.33. }
\end{figure}
As an example, in Fig. \ref{figa1c} we show the
relative energy spectra in the breakup of $^{19}$C on a $^{208}$Pb target at
67 MeV/nucleon, for a set of different binding energies of $^{19}$C : $S_n$ = 160 keV (solid), 350 keV (dotted), 
and 530 keV (dashed). In fact, there is
still an open debate  over the precise experimental determination of one neutron separation energy of $^{19}$C.
While direct mass measurements suggest a rather low value of 0.16 $\pm$ 0.11 MeV \cite{wout,orr}, recent compilations
from Audi {\it et. al} \cite{audi} suggest a value of 0.58 $\pm$ 0.09 MeV based on indirect measurements \cite{nak2,mad}.
In Ref. \cite{nak2}, Coulomb dissociation of $^{19}$C on a $^{208}$Pb target at
67 MeV/nucleon and subsequent analyses of relative energy and angular distributions yielded a value $S_n = 0.53 \pm 0.13$ MeV for
$^{19}$C, while in Ref. \cite{mad}, breakup of $^{19}$C on a $^{9}$Be target at 60 MeV/nucleon suggested two values, 
$S_n = 0.8 \pm 0.3$ MeV and $S_n = 0.65 \pm 0.15$ MeV, based on parallel momentum measurements.

Thus by accurately measuring the strength function in the breakup of $^{19}$C and comparing it with our method one should be able to
determine the most probable range for this binding energy. However, one should also keep in mind, that not only the peak position but the
whole shape of the curve is characteristic of the given binding energy (see Eq. 3).

\begin{figure}
\centerline{{\epsfig{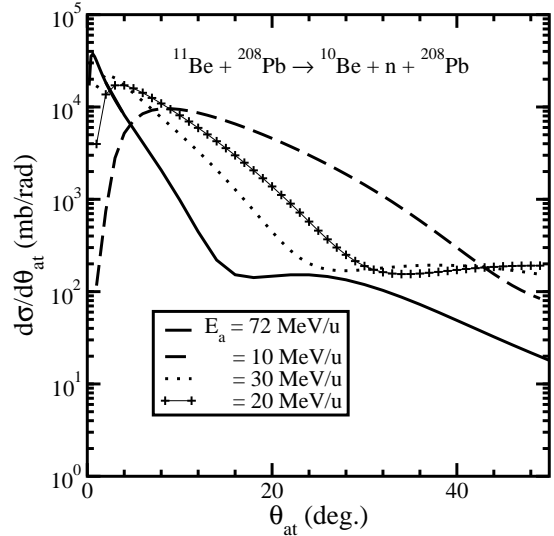}}}
%\resizebox{0.4\textwidth}{!}{%
%  \includegraphics{rel_Th.eps}}
\caption{\label{figa} Angular distribution of the projectile c.m. with respect to the target for the breakup of $^{11}$Be 
on $^{208}$Pb at different incident beam energies.}
\end{figure}
In Fig. \ref{figa}, we show the angular distribution of the
projectile c.m. with respect to the target in the breakup of $^{11}$Be on a $^{208}$Pb target 
at four different beam energies varying from low energy 10 MeV/u to 72 MeV/u.
We see that the distributions are 
more forward peaked as one increases the incident beam energy.
{%%\bf not yet mult be spec factor}

%%%%%%%%%%%%%%%%%%%%%%%%%%%%%%% E1-E2 STUFF  %%%%%%%%%%%%%%%%%%%%%%%%%%%%%%%
%%%%%%%%%%%%%%%%%%%%%%%%%%%%%%%%%%%%%%%%%%%%%%%%%%%%%%%%%%%%%%%%%%%%%%%%%%%%

\subsection{Effect of a weakly bound state near threshold}

In this subsection we investigate yet another feature of the breakup dynamics of neutron rich halo nuclei -- that 
of the relative importance of the E1 and E2 transitions to the continuum as a function of the various fictitious binding energies of the projectile. 
However, in doing so we would also like to study the effect of having a weakly bound state near the particle emission threshold,
as is the case in $^{11}$Be, whose first excited state ($p1/2$) is bound by only 0.18 MeV. 

In an ideal case one should look for a unique potential which might be able to yield the $s1/2$ (ground state) and the
$p1/2$ (first excited state) in $^{11}$Be at the right position. However this straightforward prescription fails in the 
case of $^{11}$Be, where one faces an inversion with respect to the normal ordering of states and thus in a single-particle picture 
we are forced to choose two quite distinct potentials for the ground and first excited states. In addition to the 'deep potential' used for the 
groundstate,  we need a 'shallow potential' of depth $37.38$ MeV to obtain 
the $p1/2$ bound by 0.18 MeV in $^{11}$Be.
Of course one should keep in mind that we are simulating with one-body potentials what in reality might be a more 
complicated situation (eg. a deformed core).

\begin{figure*}
%\centerline{\mbox{\psfig{file=woods.ps,clip=,width=\textwidth}}}
\begin{picture}(550,200)(0,0)
\psset{unit=1pt}
\rput(0,195){(I)}
\psline{->}(10,10)(10,190)
\psline{->}(5,110)(95,110)
\psplot{10}{90}{-100 2.18 x -50 add 4 div exp 1 add  div 110 add}
\psline{-}(10,50)(48,50)
\pspolygon[linestyle=dotted,fillstyle=crosshatch,hatchcolor=gray](10,110)(80,110)(80,180)(10,180)
\rput(32,42){2s$_{1/2}$}
\rput(45,160){\psframebox*{even}}
\rput(45,149){\psframebox*{odd}}
\rput(45,136){\psframebox*{parities}}
\pscurve{->}(70,50)(105,110)(70,150)\rput(80,75){E1}
\pscurve[linestyle=dashed]{->}(70,50)(115,110)(70,160)\rput(110,75){E2}

\rput(210,195){(II)}
\psline{->}(220,10)(220,190)
\psline{->}(215,110)(305,110)
\psplot{220}{300}{-100 2.18 x -260 add 4 div exp 1 add  div 110 add}
\psline{-}(220,50)(258,50)
\rput(242,42){2s$_{1/2}$}
\pspolygon[linestyle=dotted,fillstyle=vlines,hatchangle=45,hatchcolor=gray](220,110)(290,110)(290,180)(220,180)
\rput(255,155){\psframebox*{even}}
\rput(255,143){\psframebox*{parity}}
\pscurve[linestyle=dashed]{->}(280,50)(315,110)(280,150)\rput(290,75){E2}
\psbezier{->}(280,50)(358,92)(330,120)(358,160)\rput(345,95){E1}
\psline{->}(280,50)(358,92)
\rput(250,0){deep}
\psline{->}(360,10)(360,190)
\psline{->}(355,110)(445,110)
\psplot{360}{440}{-40 2.18 x -400 add 4 div exp 1 add  div 110 add}
\psline{-}(360,100)(405,100)
\rput(385,92){1p$_{1/2}$}
\pspolygon[linestyle=dotted,fillstyle=vlines,hatchangle=-45,hatchcolor=gray](360,110)(420,110)(420,180)(360,180)
\rput(390,155){\psframebox*{odd}}
\rput(390,143){\psframebox*{parity}}
\rput(385,0){shallow}

\end{picture}
\caption{{\label{woods}A schematic picture of potential options to generate the continuum states. In option (I) the ground state (which happens to
 be a $s$ -- state) generates both the even and odd parity continuum states. In option (II) the even and odd parity continuum states
are eigenfunctions of potentials generating the ground ($s$) and excited ($p$) states of $^{11}$Be, respectively. For more details see text.
}}
\end{figure*}
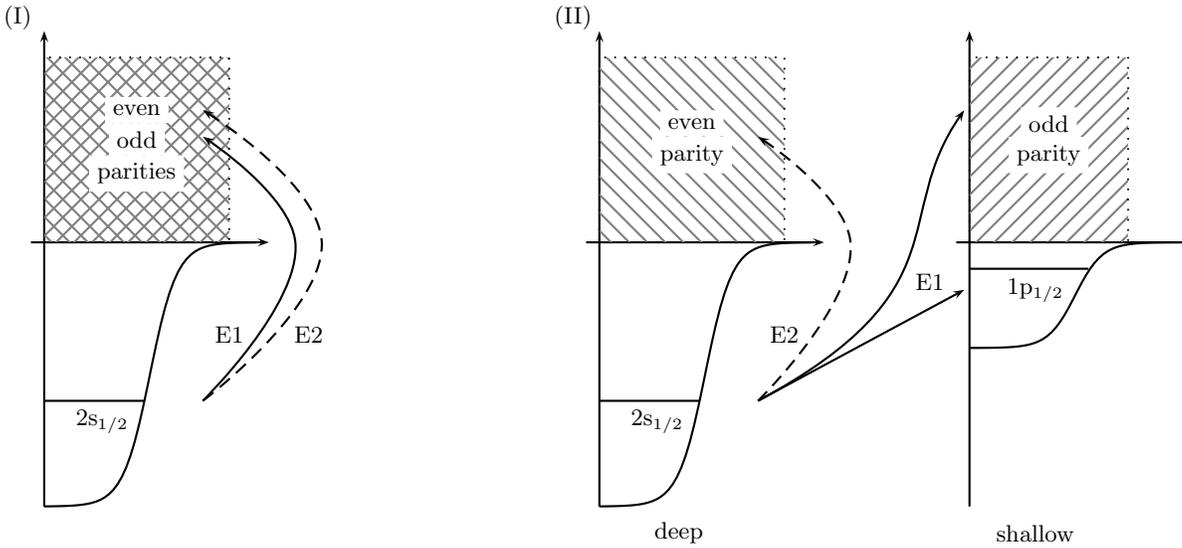 
Next comes the question of constructing the continuum states. The ground state wave function is very sensitive to the choice of the 
binding energy as we have seen in the previous sections while calculating several reaction observables. 
However, the prescription to treat the continuum states is of paramount importance. 
We thus consider two options here (see Fig. \ref{woods}) for the potential generating the continuum wave functions -- 
(I) one in which all the continuum states are generated with the deep potential, thereby ensuring that the ground state and the 
continuum state are generated by the same potential prescription and (II) another in which the even and odd parity continuum states 
are generated by the deep and shallow potentials, respectively.
Naturally, in these options the E2 transitions would remain the same as the even parity continuum states are  generated by the deep potential
in both cases. What distinguishes these options is the fact that for the E1 transitions, the relevant continuum states
are generated by deep and shallow potentials for options (I) and (II), respectively. Furthermore in option (II) the presence of the bound
$p$ state absorbs part of the E1 strength \cite{dlv}.

We shall now calculate the electric dipole and quadrupole angular distribution $(d\sigma/d\theta_{at})$ of the c.m. of the projectile in its 
breakup on a heavy target for different beam energies, within the framework of the Alder-Winther theory of Coulomb dissociation \cite{ald,aldbook}. Since the grazing angle $(\theta_{gr})$ is a function of the beam velocity, instead of directly plotting $(d\sigma/d\theta_{at})$ as a function of the projectile c.m. scattering angle $(\theta_{at})$, for each beam energy, it would be prudent to rescale the angular distribution by measuring the angles in
units of the grazing angle. Therefore we plot $d\sigma/dX$ as a function of $X=\theta_{at}/\theta_{gr}$.
Indeed the area under such a curve would again give the total one-neutron removal cross section for the concerned multipolarity and transition.

\begin{figure}
\centerline{\epsfig{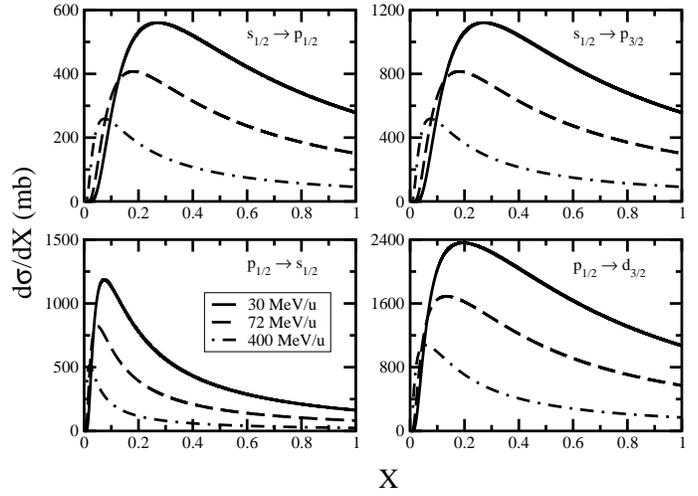}}
\caption{{\label{figE1}Scaled angular distributions for different beam energies,  from the ground and the excited states 
of $^{11}$Be for all allowed E1 transitions to the continuum are shown in the top and lower panels, respectively, in grazing angle units. 
For more details see text.
}}
\end{figure} 
In Fig. \ref{figE1}, we show the angular distributions of the center of mass of the projectile with respect to the target
for E1 in the breakup of $^{11}$Be on a $^{208}$Pb target at three different beam energies 30, 72 and 400 MeV/u as a function of $X$
for breakup from both the ground and first excited states of $^{11}$Be. For beam energies 30, 72 and 400 MeV/u, 
$\theta_{gr} = 8.12, 3.62 \textrm{ and } 0.995$ degs., respectively. The top panel shows the allowed E1 transitions from the $s1/2$ 
(ground state) to all possible continuum states, while the bottom panel shows it from the $p1/2$ (first excited state) to all possible 
continuum states.
\begin{figure}
\centerline{\epsfig{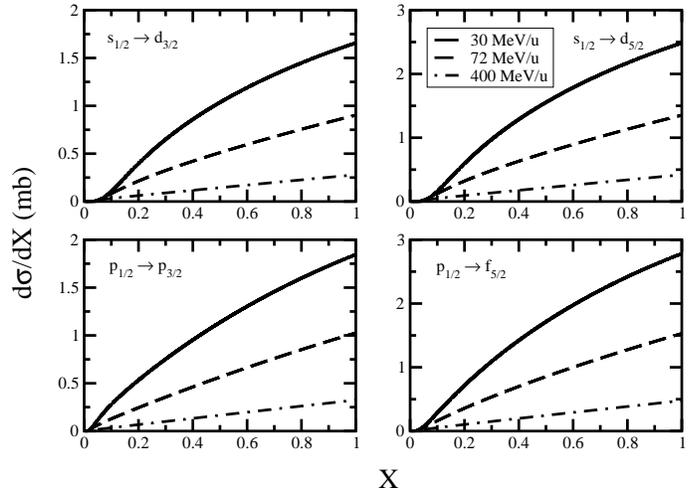}}
\caption{{\label{figE2}Scaled angular distributions for different beam energies,  from the ground and the excited states 
of $^{11}$Be for all allowed E2 transitions to the continuum are shown in the top and lower panels, respectively, in grazing angle units. 
For more details see text. 
}}
\end{figure} 

In Fig. \ref{figE2}, we show the angular distributions of the center of mass of $^{11}$Be in its breakup
on a $^{208}$Pb target for E2 transitions at three different beam energies 30, 72 and 400 MeV/u as a function of $X$
for breakup from both the ground and first excited states. The top panel shows the allowed E2 transitions from the $s1/2$ 
(ground state) to all possible continuum states, while the bottom panel shows it from the $p1/2$ (first excited state) to all possible 
continuum states.

In both these calculations the even parity continuum states are generated from the potential which also sets the $s1/2$ ground 
state (at 0.5 MeV) and the odd parity continuum states are generated within the shallow potential that also gives
the $p1/2$ excited state at 0.18 MeV. 
As expected, below the grazing angle we see a domination of E1 transitions over the E2. It is interesting to note that the 
direct breakup contribution from the first excited state is of the same order of magnitude or even greater as that from the ground state.
However, in a reaction process the breakup via the excited state is a two-step process and therefore its cross section would be
small in comparison with the direct breakup from the ground state \cite{nak3}. The inverse process, namely the capture process
via the excited state may be of importance for astrophysical considerations.
A more thorough investigation of the relative role
of the two step E1 via a weakly bound excited state with respect to a one step E2 from the ground state is called for. 

\subsection{Relative importance of dipole and quadrupole contributions to break-up cross-section}

In this subsection we investigate the relative importance of dipole and quadrupole contributions to break-up cross-section. 
This is forst done by comparing the total break-up cross section obtained within the FRDWBA, which includes the contribution 
of all multipolarities, with the pure dipole cross-section obtained within the Alder-Winther theory. 
\begin{figure}
\centerline{{\epsfig{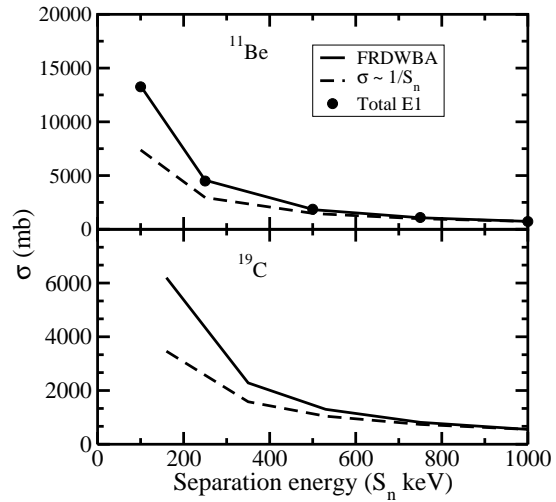}}}
%\resizebox{0.4\textwidth}{!}{%
%  \includegraphics{rel_be2_1.eps}}
\caption{\label{figb2z}Total one-neutron removal cross section as a function
of different one-neutron separation energies 
in the breakup of $^{11}$Be on $^{208}$Pb at
72 MeV/nucleon (top panel) and $^{19}$C on $^{208}$Pb at
67 MeV/nucleon (bottom panel). The solid lines show the FRDWBA values,
while the dashed lines show the $1/S_n$ curve. In the top panel filled circles show the total E1 contribution, which is
almost indistinguishable from the FRDWBA values.}
\end{figure}

As an example in Fig. \ref{figb2z}, we show the total one-neutron removal cross section in the breakup of
$^{11}$Be on a $^{208}$Pb target at 72 MeV/nucleon and $^{19}$C on $^{208}$Pb at
67 MeV/nucleon, in the top and bottom panels, respectively, as a function
of different one-neutron separation energies. FRDWBA results are shown in solid lines while the filled circles
show the total E1 contribution calculated with the Alder-Winther theory with potential option -- I (in the top panel).
Since these two results are almost coincident, one can infer that the effect of higher multipoles is negligible. 
In other words, moving away from the valley of stability towards the drip lines does not alter the predominance of 
dipole dissociation in the breakup process. Interesting to note that at higher binding energy,
the hyperbolic $1/S_n$ behaviour (dashed line in Fig. \ref{figb2z}), predicted by the simple analytical model (section 2.2)  fits quite well, 
while at low binding energy we see some deviation from the $1/S_n$ curve (see Refs. \cite{nlv,bt}).

We now turn our attention to the details of the total one-neutron removal cross section with and without taking into account the effect of the $p1/2$ excited state, within the Alder-Winther theory. 
In Fig. \ref{figcs},  the dipole and quadrupole breakup cross sections (upper panel) on Pb at 72 MeV/nucleon, and their relative 
importance (lower panel) are shown as a function of various fictitious one-neutron separation energies of $^{11}$Be. 
 The solid black lines in both figures are those in which even parity continuum states are generated from the deep potential and the odd parity continuum states are generated from the shallow one (option II above). The dotted lines show those in which a unique single particle potential is used to generate both the ground and the continuum states (option I above). It is interesting to note that, although the absolute cross sections are strongly dependent on the binding energy, the relative importance 
of E1/E2 transitions shows a much weaker behaviour in both options (and especially in option II).

\begin{figure}
% Use the relevant command for your figure-insertion program
% to insert the figure file. See example above.
% If not, use
%\resizebox{0.75\textwidth}{!}{%
%  \includegraphics{E1byE2.eps}
\centerline{\epsfig{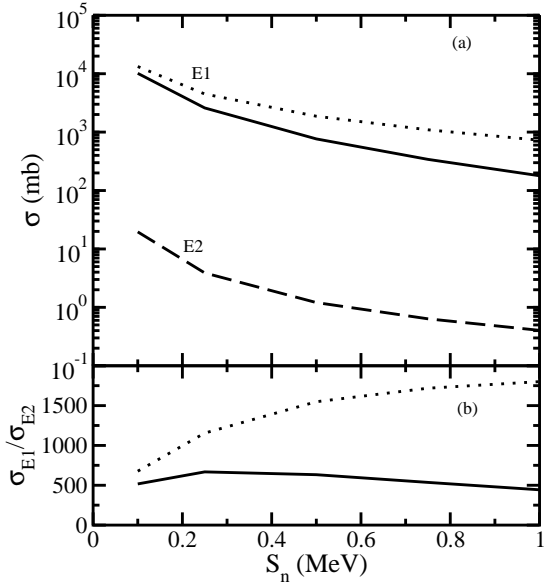}}
%\includegraphics[width=.75\textwidth]{E1byE2.eps}
%\vspace*{5cm}       % Give the correct figure height in cm
\caption{\label{figcs}(a) Total and  (b) relative importance  of dipole (solid) and quadrupole (dashed) breakup cross sections on Pb at 72 MeV/nucleon, as a function of one neutron separation energies of $^{11}$Be. The solid lines in both figures are those in which even parity continuum states are generated from the $s_{1/2}$ and the odd parity continuum states are generated from the $p_{1/2}$ -- potential option II. The dotted lines show those in which a unique single particle potential is used to generate both the ground and the continuum states -- potential option I.}
       % Give a unique label
\end{figure}

%%%%%%%%%%%%%%%%%%%%%%%%%%%%%%% CONCLUSIONS  %%%%%%%%%%%%%%%%%%%%%%%%%%%%%%%
%%%%%%%%%%%%%%%%%%%%%%%%%%%%%%%%%%%%%%%%%%%%%%%%%%%%%%%%%%%%%%%%%%%%%%%%%%%%

\section{Summary and Conclusions}
In this paper, we have compared two different theoretical models of breakup reactions by calculating several reaction observables like  
relative energy spectra, angular distributions and breakup cross sections, taking the neutron separation energy as a parameter.
Thus we have theoretically simulated situations ranging from weakly bound isotopes to stable ones for the same angular momentum configuration 
of the system.

We have first calculated the relative energy spectra in Coulomb induced breakup processes for various projectiles 
and beam energies as a function of the   
neutron separation energy, within the framework of the post form FRDWBA. In this model the electromagnetic 
 interaction between the core and the
 target nucleus is included to all orders and 
 the breakup contributions from the
entire continuum corresponding to all the multipoles 
and the relative orbital angular momenta between the valence
nucleon and the core fragment are included in the theory.

We have studied the relative importance of dipole and quadrupole breakup contributions in Coulomb dissociation
under the framework of the Alder-Winther theory.
We have constructed a single particle toy model for the structure of the halo nucleus $^{11}$Be, with
separate Woods-Saxon potentials describing the bound states of the system, namely $s1/2$ (ground state) and 
$p1/2$ (first excited state). To the unique
continuum generated for each of these configurations,  
we calculated Coulomb 
breakup cross sections on Pb by varying again the one neutron separation energy of $^{11}$Be. 
We have found that the relative importance of the E1/E2 transition remains fairly constant.

Finally the investigation of the one-neutron breakup cross section as a function of separation
energy obtained by comparing the results of two theories revealed that, as one goes away from the valley of stability
towards the drip lines  (where one would encounter predominantly weakly bound isotopes) higher
multipoles, other than the dipole, do not play any significant role in the breakup process. That the two results were almost identical
also opens up an interesting opportunity. In the calculation of astrophysical $S$ - factor via the Coulomb dissociation 
method it is crucially important that only a particular multipolarity is almost solely responsible for the breakup process.
A word of caution is in order here: these results have been obtained for a weakly-bound neutron halo and in principle one might expect 
a different behaviour for the case of heavier/charged clusters.

%%%%%%%%%%%%%%%%%%%%%%%%%%%%%%% REFERENCES  %%%%%%%%%%%%%%%%%%%%%%%%%%%%%%%%
%%%%%%%%%%%%%%%%%%%%%%%%%%%%%%%%%%%%%%%%%%%%%%%%%%%%%%%%%%%%%%%%%%%%%%%%%%%%

\end{document}